# FACTORS INFLUENCING RISK ACCEPTANCE OF CLOUD COMPUTING SERVICES IN THE UK GOVERNMENT

Gianfranco Elena[1] and Christopher W. Johnson[1]

[1]Computing Science, University of Glasgow, Glasgow, UK

## ABSTRACT

*Cloud Computing services are increasingly being made available by the UK Government through the Government digital marketplace to reduce costs and improve IT efficiency; however, little is known about factors influencing the decision making process to adopt cloud services within the UK Government. This research aims to develop a theoretical framework to understand risk perception and risk acceptance of cloud computing services.*

*Study's subjects (N=24) were recruited from three UK Government organizations to attend a semi structured interview. Transcribed texts were analyzed using the approach termed interpretive phenomenological analysis. Results showed that the most important factors influencing risk acceptance of cloud services are: perceived benefits and opportunities, organization's risk culture and perceived risks. We focused on perceived risks and perceived security concerns. Based on these results, we suggest a number of implications for risk managers, policy makers and cloud service providers.*

## 1. INTRODUCTION

Cloud computing represents a new model to create and distribute scalable software and infrastructure services.While cloud services such as Dropbox, Gmail or Netflix have been used by millions of individuals, it is quite recent that Government organizations have begun to use cloud services as a solution for their IT needs.

In 2011, the UK Government published the "Government Cloud Strategy" to promote the adoption of Cloud services as a way to improve the cost efficiency, flexibility and interoperability of the IT services. Until today, an online catalogue of 13,000 cloud services, including email, enterprise resource planning, learning management, office productivity, polls/surveys and analytics, is available on the Government digital marketplace[1].

Despite the adoption of the "Cloud First policy", the adoption of Government cloud services still represent just 1% of the central UK Government ICT bill. Little is known about factors that influence risk acceptance of cloud computing services within the UK Government organizations.

The rational decision-making process for evaluating cloud computing risks has to take into consideration the actual risk or "objective" risk and the "perceived" risk, based on the judgements of those assessing the risk [2]. It is hard to distinguish between "objective" risk and "perceived" risk of cloud computing. For relatively new technology like cloud services, objective measures of the risks are very hard to achieve and "objective" risks must be predicted by using complex techniques which usually requires subjective judgments of experts [3]. The difference between "objective" risks, estimated by technical experts, and public "perceived" risks can create some difficulties for decision and policy makers. In fact, if public "perceived" risk is higher than "objective" risk, it becomes a challenge to adopt new digital innovations [4].

Understanding how risk perception and risk attitude influence risk acceptance of cloud services within Government departments may help risk managers and policy makers to prevent either that an overly cautious risk culture results in a failure to seize important opportunities or taking exaggerated risk without regard to the potential risk. In that regard, this study represents a contribution to support actions that aim to align risk exposure to risk appetite in order to maximize the efficiency and improve business services innovation taking acceptable risks.

This study aims to draw a map of the factors that are likely to influence the adoption of cloud computing services [5]. By using grounded theory analysis [6], a descriptive theoretical framework of the factors influencing risk acceptance of cloud services is formulated. We used a semi-structured interview to examine Government professionals' behavioral intention formation when assessing cloud computing services risks and benefits.

The aim of this research, therefore, is to explore the following issues:

- Which factors influence risk acceptance of Cloud computing services?
- What are the most important perceived risks of Cloud computing services?

This paper is structured as follows: first, we revise the state of the art for adoption of Cloud services in Government organizations. Second, we review the literature on organization's risk culture, and perceived benefits and risks of cloud software as a service. Third, we present our methodology, based on grounded theory, to investigate the opportunity-risk framework for the adoption of Cloud services within Government organizations. Next, we present our empirical analysis of the results, and we conclude with a discussion of our findings, the theoretical and practical contributions of our work, its shortcomings, and future research directions.

## 2. CLOUD COMPUTING ADOPTION IN GOVERNMENT ORGANIZATIONS

Recently, many EU countries have developed a Cloud national strategy,according to the recommendationsof the European Commission's Cloud Strategy, but only a few of them have developed a Governmental Cloud infrastructureto support the public administration. Previous to the study of the Cloud risk acceptance framework, it is important to understand the state of cloud adoption inthe European Member States and the main benefits and concerns for the adoption of Government Cloud services.

According to Z wattendorfer [7], eight European countries have already planned to use Cloud Computing (see Table 1). Despite three countries (UK, Spain, and Denmark) have already implemented a Cloud infrastructure, the full implementation of their national Cloud Computing strategy will still take another few years [8].

Table 1. Comparison of Cloud computing adoption in e-Government across eight European countries made in [7]

| Country | National Strategy | Cloud Adoption | Deployment Models | Cloud Services | Examples of Cloud services |
|---|---|---|---|---|---|
| Austria | Yes | Planned | Public Private Community | IaaS PaaS SaaS | Backup Collaboration Services Identity as a service |
| Denmark | No | Planned Executional | Public Private Community | SaaS | Email Procurement |
| Finland | No | Planned | | | |
| France | Yes | Development | Community | IaaS | |
| Germany | Yes | Planned | | | |
| Ireland | Yes | Planned | Public Private Community | IaaS PaaS SaaS | Open Data Collaboration Services Email |
| Spain | No | Planned Executional | Public Private Community Hybrid | IaaS PaaS SaaS | Open Outsourcing Email Storage/Backup Office Collaboration |
| UK | Yes | Planned Development | Private Community | IaaS PaaS SaaS | Email Office CRM |

The adoption of Cloud Computing in the Governmental organizations offers many potential benefits. First, the savings obtained from operating and maintaining their hardware and software infrastructures [9]. Second, an increased capability to test and procure IT capacities that they may not have been able to afford in the past [10]. Third, the flexibility to manage IT resources allows scaling up and down capacity on demand and only pay for the actual usage. Also, cloud platforms enable to use an agile development environment that makes it easier for IT professionals to develop applications quickly and to adopt them instantly[11][12].

On the other hand, recent studies have grouped the cloud risks into four categories [13]: policy and organizational risks (e.g. data lock-in, loss of governance), technical risks (e.g. cyber-attacks, loss of data), legal risks (e.g. data protection and legal jurisdiction), and other risks (e.g. network problems, internet connection).

First, theinteroperability of different cloud platforms it is still hard to achieve. A lack of standardization means that government organizations would not be able to shareeasily information with other organizations as well as transfer their data from a cloud service providerto another [14].

Second, security and privacy issues are considered as key factors for the adoption of cloud services [15]. The main security challenges are data protection and compliance, identity and access management, auditing, as well as risk management and detailed security SLA formalization [16].

Third, government organizations have concerns about privacy and data confidentiality a lack of control over the physical infrastructure [14][17] and for the IT performance which it is controlled not by their staff but by off-premises cloud providers; and that they may not be able to make necessary changes in application features easily and when needed[18][19].

Fourth, there is a concern about service availability and reliability especially regarding the unexpected cloud system downtime and disruption [9].

In summary, cloud computing offers many advantages and challenges for government organizations [19]. Some of these challenges are technical while some are related to the uncertainties derived from engaging with a recent innovation. An objective of this study was to identify perceived factors that may discourage IT government professionals from adopting cloud computing [20].

## 3. THEORETICAL FRAMEWORK

### 3.1 Organization's Risk Culture

"Organization's risk culture consists of the norms and traditions of behaviour of individuals and of groups within an organization that determine the way  in which they identify, understand, discuss and act on the risk the  organization confronts and takes" [21]. Many factors influencing risk-related behaviours have been studied by psychologists and sociologists. Main components include risk perception and risk propensity [22], decision-making process [23] and personal characteristics of risk takers [24][25].

The question if public risk managers are more risk averse than private risk managers has been debated in the last decades [26]. It is a common view that public risk managers have little incentive to take risks and that risk aversion could undermine effective decision making process. Despite many studies investigated the differences between risk aversion of private and public risk managers and the effect of risk aversion on managerial decisions [24][25], none of them proved systematically differences  between public and private organizations [27][22]. We identified some reasons that could influence risk aversion in public risk managers and decision makers. First, the lack proprietary property of rightsis a disincentive to take managerial risky decisions in the public sector. Second, public risk manager are due to the stringent supervision of their decisions and risk taking behaviour could be hard to explain in the case of poor results. Third, the bureaucratic process and the higher degree of formalism to change the "status quo"introducingnew processes and digital innovation can be discouraging [28][29][30].

Since little is known about risk aversion in public organizations and their managers about the adoption of cloud computing services, it would be useful to have a better understanding of risk taking by public managers.

Given the importance of these issues, this research investigates factors influencing public managers' risk-taking about the adoption of digital innovation like cloud computing services. Government organizations may have different ranges of risk attitudes towards different risks. Perceived risks that are acceptable for one department could be not acceptable for another one. Risk attitude depends on the perceived opportunities gained in comparison to the related potential losses.

### 3.2 Perceived Benefit and Risk

The theory of riskdefinessix main dimensions of perceived risk: performance, financial, time, psychological, social, privacy [5][31][32][33][34].Consumers perceptions of risks involved in the adoption of SaaS services have been studied in the past years [35][36][37] (Table 1).

On one hand, the main benefits are the pay-per-usage, end-user convenience and ease in installing and managing software, improved software quality. On the other hand, the most frequently mentioned risks are data security and system integration with the legacy system.

Table2. Benefits and risks of Cloud services adoption.This table is based on a meta-analysis of[35]

| Authors | Perceived Benefit | | Perceived Risk | |
|---|---|---|---|---|
| | Research content | Main result | Research content | Main result |
| [35] | Investigated the perceived benefits of SaaS | - Cost advantages<br>- Strategic flexibility<br>- Focus on core competencies<br>- Access to specialized resources<br>- Quality improvements | Investigated the perceived risks of SaaS | - Performance risk<br>- Economic risk<br>- Strategic risk<br>- Security risk<br>- Managerial risk |
| [36] | Explored perceived benefits from the perspective of (SaaS) customers | - Pay only for what is used<br>- Easy and fast deployment to end users<br>- Monthly payments<br>- Encourages standard systems<br>- Requires fewer in-house IT staff members and lower costs<br>- Always offers latest functions | Explored the perceived risks from the perspective of (SaaS) customers | - Data locality and security<br>- Network and web application security<br>- Data integrity and segregation<br>- Authentication and authorization<br>- Virtualization vulnerability<br>- Data access and backup |
| [37] | Examined the benefits of deploying cloud-based systems (SaaS) | - No installation and maintenance of software<br>- No software expertise necessary<br>- Eliminates the need for an ICT department<br>- No complicated license management<br>- Access to software without a need for upfront investments | Examined the problems associated with deploying cloud-based systems (SaaS) | - Need for contractual expertise<br>- Quality assurance<br>- Ensuring the accountability of service providers<br>- Problems shift to composing and its integration with legacy systems<br>- Assurance that data are backed up and can be recovered |

The analysis of this qualitative study was based on the review of factors influencing risk acceptance and risk attitude of cloud services. We used these factors, in the post interview, as part of the conceptual framework to derive our themes and the subsequent analysis.

## 4. METHODOLOGY

The aim of this research is to explore the factors influencing the risk acceptance of cloud computing services in the UK Government organizations. We conducted a number of semi-structured interviews to collect relevant information. In that regard, we paid attention to select the relevant sources of information, avoid indications of causal relationships, define variables for building a model, and take into consideration contextual factors like organizational risk attitudes and culture. Transcribed texts of the interviews were then analyzed and coded according to the grounded theory methodology [38]. Grounded theory analysis was successfully used in other similar studies for generating ideas' frameworks, not connected a priori to pre-existing theories, by empirical data. In this sense, "using the grounded theory methodological framework it is possible to ensure that the theoretical description appropriately reflects the empirical setting and that the theoretical framework is actually generated by the data description"[39].

### 4.1 Data collection

A pilot study on four individuals was used to test the interview process. Study subjects were then recruited from three UK Government organizations. An email providing background information of the research was sent to all participants through their chief division, seeking voluntary participation in the research. Potential participants who expressed interest in the research were provided additional information through email in the form of an information sheet providing background information about the planned interviews. Upon confirmation to assist with the research, a mutually suitable time was arranged to conduct the interview. Twenty-four Government professionals (twenty males) took part in this study between January and March 2014 (named P1,…, P24). Subjects were unknown to the researcher before the interview. All interviews were digitally recorded and on average each interview took approximately thirty minutes. The number of participants was deemed satisfactory according to "theoretical saturation" principle [40].Data was collected through semi-structured interviews that consisted of a series of open-ended questions. The interview was structured in four main parts. First, we collected general information on the participant and his role in the organization. Second, we elicited predictors of intention to adopt cloud-services as seen by the professionals working in government organizations. We aimed to obtain insight into their views about benefits and barriers of adopting a cloud computing technology. Specifically, question one reads "If handled digital innovation within your organization, what would influence your decision to adopt cloud services and why?" The second question reads "What are the benefits and barriers that you would consider about adopting cloud services in your organization?". Third, we investigate the perceived risk of cloud-services. Question three reads "What are the most important perceived risks about using cloud computing services in your organization?", Question four reads "Can you describe these risks and explain why they are so important?". In the end, as a result of the first two iterations, we focused on "security risk" by asking "How important is security risk about using cloud services within your organization and why?". Question six reads "What are the most important security risks for adopting cloud services and why?".

Eighteen of the interviewees had a managerial role within the organization. The average work experience was 18 years (SD 6.8), and the average age was 44 years (SD=7.8).

### 4.2 Data Analysis

Transcribed texts were analyzed "using the approach termed interpretive phenomenological analysis" [41] which aims to understand the participant's point of view by interpreting his answers. First, all the interviews were printed out and carefully examined to make sense of the general content and meaning of the texts. Then, we conducted two main activities: open coding for the identification and labelling of concepts and axial coding for the definition of the relationships between concepts. The qualitative data handling program Atlas (Visual Qualitative Data Analysis-Management-Model Building) was used to assist analysis [42]. Data were analyzed and discussed by two researchers, "taking notes and writing memos along the lines" [38]. Open and axial coding generated a list of main categories (i.e. families of concepts) and concepts which allowed drawing a network of concepts and categories. The outcome of both data collection and data analysis was validated. In the rest of the paper selected outcomes from this empirical study are presented. The process is described in figure 1.

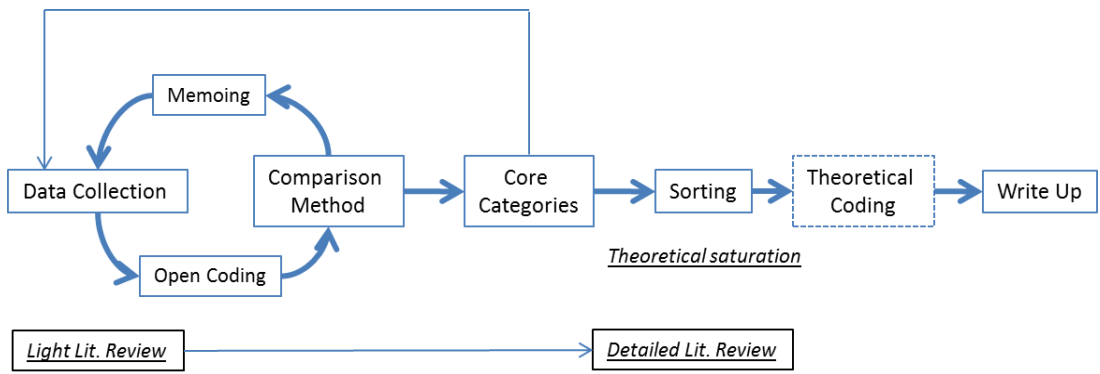

Figure 1.Transcribed texts analyzed using the "termed interpretive phenomenological analysis".

## 5. RESULTS

The result of the grounded theory analysis is basically a list of concepts grouped into categories and connected by hierarchical levels. The most significant networks of concepts generated through the empirical analysis of collected information were:  a. Factors influencing adoption of cloud services; b. Perceived risks of adopting cloud services; c. Perceived security risks.  These three core areas will now be shown in some detail focusing on the most important concepts/properties, where the importance is expressed in  terms of so-called ''groundedness'', i.e. the number of citations (the number of times  a concept/property was mentioned during interviews) [38].In order to analyze participants' responses, the data were coded in themes (i.e. concepts, properties) and grouped into categories based on underlying principles. In total, 331 coding themes and 22 concepts were formed and labeled as described in figure 2. These categories were created based on words or phrases used by respondents, which were attributed to the object under examination.

The responses to the six questions were summarized and described below. As the format of the study involved open-ended questions, this often resulted in participants providing more than one response to each question. As a result, the percentage value indicates the importance of the network connection based on the number of times a concept/property was mentioned during interviews [43].

### 5.1 Reliability

The results appeared reliable, as themes from the pilot study and both organizations were similar, and an independent assessment of pilot transcripts revealed consistency in derived concepts [38]. Agreement rate amongst the three specialist independent raters was 92%.

## 5.2 Validity

### Data triangulation

This study employed data triangulation, in that data were collected from two differing organizations and professionals with a reasonably wide experience background. The data did not significantly vary between the organizations or participants [43].

### Respondent validation

Respondent validation involved the comparison of the investigator's account with the pilot subjects' answers. The level of concordance was tested by asking participants if the interpretation by the researcher was accurate in terms of what they were trying to communicate. This was carried out using a linear score where 10 represented total agreement and zero represented no agreement [6]. The mean scores ranged from 7.5 to 8.6 for the pilot subjects. This suggests a high degree of respondent validity.

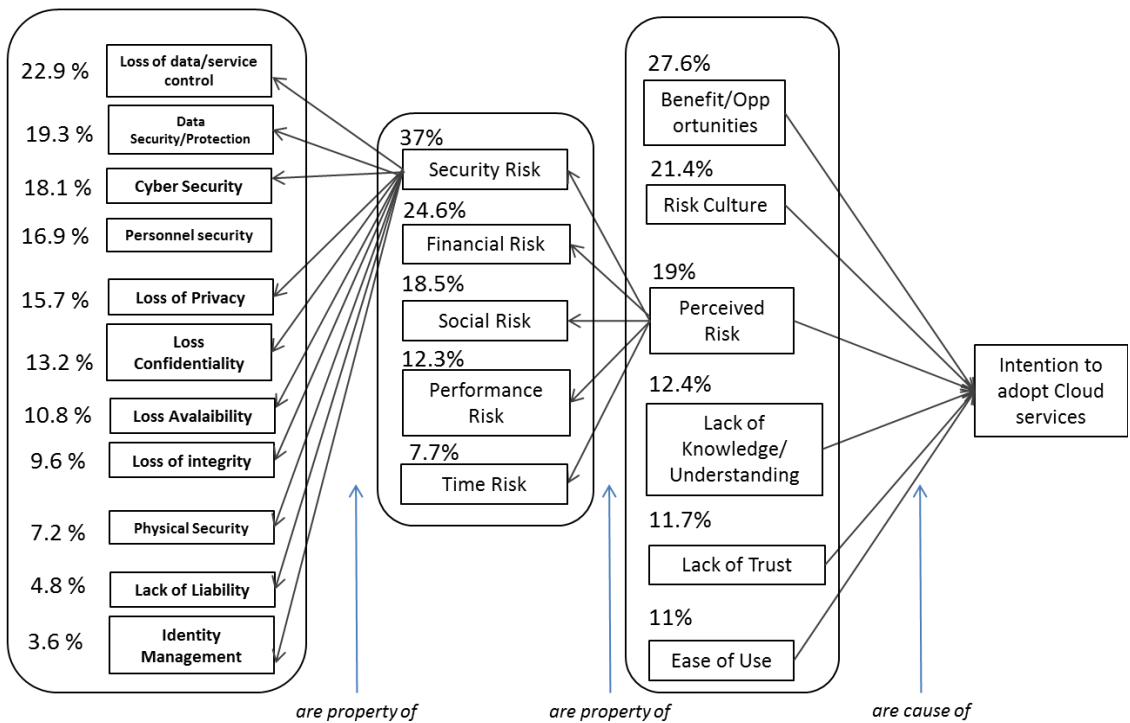

Figure 2.N=24. Description of the three families of concepts and their relationships.

## 6. DISCUSSION

Previous studies explored factors affecting the adoption of cloud computing services, including the technical, process, and economic factors [44] as well as perceived risks and perceived benefits [37][45]. But little is known about the factors that inhibit or enable the adoption of cloud services in Government organizations. The primary goal of this study is to gain a deeper understanding of the components that influence risk acceptance of cloud computing services in the UK

Government. In that regard, we used a grounded theory approach [43]to elicit factors influencing cloud computing adoption and build a theoretical framework generated from the data description. We used qualitative analysis to interpret the data applying the principle of theoretical saturation to discover concepts and linkages between concepts [6]. Based on the results of the interpretative data analysis we conducted a literature review to define better derived concepts and families of concepts.

This in-depth interview study highlights that the most important factors influencing risk acceptance of cloud computing services within the UK Government organizations are: perceived benefits/opportunities, organization's risk culture, perceived risks, lack of knowledge/understanding, lack of trust and ease of use (see figure 2). Risk concern was a consistent issue raised by all professionals without prompting. Our findings about perceived risks are in line with the previous studies[32][5][33][34][31] and indicated the following perceived risk facets: security risk, economic risk, social/reputational risk, performance risk and time risk. In particular, security risk resulted to be the main concern. Participants reported a higher level of risk perception for the possible loss of data control and data protection, lack of cyber security and personnel security, loss of privacy and confidentiality, loss of availability and integrity.

Cost savings [44][45] on software procurement and maintenance are usually one of the strongest determinants to start using cloud services in the private sector.An interesting finding about perceived benefits of cloud services in Government organizations was that our participants focused on the importance to increase the flexibility and the interoperability of the IT services provided: *"I believe that cloud services could help to easily share information within the organization and with citizens" P1; "It should reduce the time to procure and upgrade software application services", P5; "It will facilitate information access from multiple locations and devices", "it can improve interoperability with other organizations", P8; "it makes easier to collaborate and exchange data with other departments"* , *P10*. At the same time, participants confirmed that cost advantages play an important role in the decision to adopt cloud services: *"it can improve economies of scale and produce relevant cost savings", P3; "it could reduce the need for multiple data centers and software maintenance", P9; "it should help to have fewer data centers saving on hardware and energy costs", P13; "it can save time and improve organization's productivity", P19.*This result suggests that Government organizations primarily need to be more efficient by improving their interoperability and flexibility. There is a need to collect, share, visualize, analyze, store and retrieve information more efficiently. With thousands of different IT systems and services available, it is a challenge to collaborate easily and share information with other organizations and citizens.

Organization's risk culturehas an influence on how government professionals identify, understand, discuss and act on the risk their organization confronts and takes [21]. In that regard, participants indicated that cloud service adoption could be influenced by the following aspects:*"Government organizations must follow their IT security policy which specifies how to procure and use available IT services", P14; "a top-bottom approach is used to define what kind of IT services are necessary and should be available to end users. End users can represent their needs or suggest upgrades, but they are rarely implemented…", P18;"there is a hierarchical structure for taking decisions but often the top management is not aware of the needs of the end users…", P6; "I feel a different sense of responsibility for taking personal decisions or decisions affecting the Government organization in which I work…", P8; "I have little or no incentive to take risks on my work. If something wrong happens it would be hard to justify and motivate risky choices", P17; "it is easier to take risky decisions that are in line with the objectives fixed by the top management", P21.* These results suggest that Government professionals are more risk averse than private risk managers [26] and more reluctant to adopt cloud services within their work environment. First, they have little interest to introduce cloud services especially if there is no clear communicationof the senior management to enhance digital innovation.Participants suggested that they usually trust and follow the directives they receive from their security department. A second reason is that the hierarchical structure of government organizations usually requiresa long bureaucratic process to approve the use of new technology or procedure. Many risky decisions are not taken just because it is easier and quicker to go with the old solution [29][30].

In line with other studies about perceived risks[5][31], our findings confirmed that most important perceived risks of cloud services are: social/reputational risk,financial risk, and security risk. Participants defined social risk as: *"social risk for using cloud services is very high because of the potential damage and loss of reputation in case of leakage of personal data and unavailability of the cloud services", P12; "...the negative impact on public opinion in case of cyber incidents could result in a lack of trust in government authorities", P9; "...as a risk manager I avoid taking risky decisions that could have a bad impact on the image of the organization...", P19.* Results suggest that social risk is perceived as very high especially in consideration of the effects of the social amplification phenomenon caused by media and social media coverage in case of cyber incidents [46].

Also, participants defined the financial risk as: *"I wonder if using cloud services will replace legacy IT systems or just add something else to what we have already", P12; "...sometimes we are locked into existing contract that will last for years. It is not easy or convenient to replace these contracts... financial penalties from withdrawing could be applied...", P23; "cloud services need to prove efficiency and effectiveness before spending money on new IT systems...", P24; "High legal fees to personalise terms and conditions of use should be taken into consideration to evaluate the economic convenience of using cloud services...", P14; "...Costs higher than expected shouldn't be a bad surprise...", P20.* It is clear that cost savings are an important perceived benefit. Government professionals are attracted by the economic convenience of using cloud services but are not confident about the total cost of replacing their legacy systems with cloud services. They are worried that hidden costs would affect the economic convenience of their decision.

Finally, participants described security risk as the most important concern to adopt cloud services. A number of factorsrelated to security risk were described as follows:

Loss of service control: *"...our technical staffs administer and have full control onall IT systems. It is hard to imagine what could happen if something wrong happens with cloud services as we need to have quick answers and solutions...",P2; "...cloud service providers have full control of the cloud infrastructure and could change the rules without our consent or enforce new terms of use", P14.*

Data security and protection: *"how can we know how cloud service providers will store and protect our data if we use cloud services?..." P3; "...will cloud service provider provide a report of major and minor data security incidents? It is important to know what happens in real time in order to react appropriately...",P22; "what is the level of data protection provided for data stored in the cloud data centre? What are the data security standards for governmental clouds?...", P21; "...are data stored with adequate level of protection and secured from the data centre to the user?", P15; "...UK law is very cautious about personal data and there are legal issues if data is not stored in the UK. It is important to know where data are stored and protected...", P19.*

Cybersecurity: *"...using cloud services could improve the risk of cyber-attacks. If data are concentrated in one place it could be easier to attack...", P11; "...such a concentration of sensitive data in the same place could trigger lots of interest and potential threats", P15; "protecting a government cloud infrastructure from cyber-attacks is a bigger challenge...what is the responsibility of the end users and the cloud service provider?, P18.*

Personnel security: *"...how do we know who has access to the data centre infrastructure? What happens if malicious insiders have access to the data?",P4; "...what kind of controls, policies, standards, security procedures are in place in the cloud service provider...", P5; "...How cloud service provider can avoid leakage of information? How can they ensure that their personnel respects the highest security standards?", P15.*

Loss of privacy: *"...every year many portable devices are stolen or lost.... using cloud services in a government organization could improve problems of privacy...", P7; "with so many personal data there is a risk that unauthorized persons could access personal information of ....", P24.*

Loss of confidentiality and integrity: *"...many data exchanged within and outside the UK government are to be considered sensitive and, using cloud services, the risk of losing data confidentiality and integrity could improve...", P19.*

Loss of availability: *"...there are many examples of cloud services that were not available for days due to technical problems. Services provided by the Government shouldalways be available...I am not sure that using cloud services can improve resilience of IT services...", P5; "...if the internet connection is not available than cloud services will not work...", P3; "there are many examples of incidents that made unavailable cloud services for the day. Since Government critical infrastructures need to be extremely resilient, it is a challenge to plan the adoption of cloud computing services within a Government organization...", P18.*

These results suggest that security concerns are one of the most important perceived risks for using cloud computing services in Government organizations. There could be many reasons for this common risk perception. First, there is a need to increase the level of trust in the quality of service provided by Cloud service providers. Second, term and conditions of the Service Level Agreements are not known or unclear. Third, Government Cloud services available through the UK Cloud store were unknown to the majority of the participants. Fourth, it is unclear how cloud service providers willadequately protect the privacy and confidentiality of data stored in their cloud infrastructures. Fifth, it is hard, if not impossible, to supervise the security activities performed by the personnel of the cloud service providers.

In summary, this study contributes to build a conceptual framework that describes the linkages between different factors that influence risk acceptance of adopting cloud computing services in Government organizations. We focused on main perceived risks discovering the importance of security risk and its contributing factors.

# 7. CONCLUSIONS AND FUTURE RESEARCH

Based on the previous considerations we suggest someactions that could support the adoption of cloud computing services in Government organizations.

## Implications for National and Member Statespolicy makers

- Cloud Computing strategy should be part of the national ICT strategy, enabling central Government to quickly procure and deliver digital service innovations to citizens;

- Cloud Computing strategy should define clear objectives, timetables and managers' responsibilities within each Government organization taking into account the implications for security;
- Should promote a centralized Governmental cloud store to procure all IT services across Government organizations;
- Provide case studies of successful implementation detailing pros cons and total costs
- Should define a plan of incentives for Government risk managers and decision makers who are responsible to introduce digital innovation
- Should assess the results achieved and lesson learned annually;
- Should define a generic security framework for governmental clouds [47];
- Manage appropriately risk communication in case of cyber incidents to avoid the effects of the social amplification factor [48];

**Implications for Government' risk managers and IT decision makers**

- Should definedesired risk exposure and risk tolerance for using cloud services within different organizations
- Should improve services efficiency and innovation by taking acceptable risks
- Should prioritize the procurement of cloud services to enhance IT services interoperability inter and intra organizations
- Should effectively communicate the benefits and opportunities provided by new cloud services
- Should define clear objectives and managerial responsibilities to achieve adoption of cloud computing services
- Should enable end users to learn and use efficiently the new cloud services

**Implications for Cloud Service Providers**

- Should provide clear information about Service Level Agreements, specifying level of responsibility and time of intervention;
- Should provide a flexible pricing plan that enables government organizations to pay as per use;
- Should provide an Incident Response service to enable customers identify and react quickly to potential cyber-attacks;
- Should protect consumer data against physical tampering, loss, damage or seizure;
- Should ensure that all the staff be subject to personnel security screening and security education for their role;
- Should provide consumers with the tools required to enable them securely manage their service;
- Should allowaccess to all service interfaces only to authenticated and authorised individuals.
- Should provide secure service administration mitigating any risk of exploitation that could undermine the security of the service;
- Should provideconsumers with the audit records they need to monitor access to their service and the data held within it.

The present research aimed to identify factors influencing the adoption of cloud computing services within Government organizations. The results revealed six main drivers that require to be interpreted in context. Specifically, the selected sample originated from only two UK Government organizations, which bears the risk that their views on cloud computing adoption could be biased by a common previous experience with legacy IT systems. While there is no evidence to suggest that this sample was not representative of the wider Government organizations population, future research should include samples from various Government organizations to confirm the validity of the current findings.

Additionally, [49] it needs to be remembered that the sample size included 24 participants. From both a theoretical and applied perspective, future research should also investigate the applicability and predictive validity of the new key variables identified in the present research. Moreover, future studies should test if perceived risk can be determined by the identified variables, and how important is the perceived risk to predict the risk acceptance for adopting cloud computing services. For future research it would be interesting to evaluate the statistical significance of the identified variables and the correlation between independent and dependent variables. In particular this will enable to define a high level algorithmto predict risk acceptance of Cloud computing services in Government organizations.

## Appendix 1

| CONCEPTS | PROPERTIES (from Open coding) / NUMBER OF TIMES the property was mentioned | % |
|---|---|---|
| **PART 1.** 1. "If you were responsible for digital innovation within your organization, what would influence your decision to adopt cloud services and why?" / 2. "What are the benefits and barriers that you would consider about adopting cloud services in your organization? | | |
| **Benefit/Opport unities** | Share information within the organization and with citizens(9), Cost savings (3), Economies of scale (2), 24/7 support (2), reliability (2), time to procure and implement new sw (2), easy sw upgrade (2), access from multiple locations and devices (3), interoperability with other organizations (3), standardization of software and data (2), improving resilience (2), Data centres more efficient (1), Compatibility (1), Easier maintenance (1), collaborative functionality (2), more functionality (1), improve productivity (1), improve accessibility (1) | 27,6 |
| **Risk Culture** | public sector employees are resistant/not willing to change (2), reluctance to innovation(1), our employees are risk averse(4), there is no incentive to take risks(2), G-Cloud is not proved(1), top management is not aware/able to weight latest technologies risk opportunities (2), Government is not an early adopter/prefers to adopt tested and mature products (6), employees are reluctant to share information(1), IT Departments take decisions based on risk assessment and Cost Benefit Analysis (1), information security is devolved to Government(1), there are many different organizations with different needs and level of IT knowledge (1), complex and bureaucratic process to change (1), influence of policy, rules and procedures (8) | 21.4 |
| **Perceived Risks** | Loss of control (8), privacy (5), cyber security(3), cybercrime (3),Need to maintain the legacy systems (3), Locking into existing contracts (1), Reputation could be damaged in case of security incidents (3), Cloud services could not achieve expectations(1), Time to understand terms and conditions of use (1) | 19 |
| **Lack of Knowledge/ Understanding** | lack of knowledge (6), lack of knowledge of Government plans and ambiguity about G-Cloud (3), lack of understanding of security issues (3), not clear the need for cloud services (2), cloud term is confusing (1), lack of awareness how to access cloud services (1), fear of unknown (1), lack of awareness of benefits and services available (1) | 12.4 |
| **Lack of Trust in CSP** | lack of trust in CSP (4), concerns about most important Cloud Service Providers (4), CSP pursue their commercial interests which are not the same of Government (1), CSP providing private cloud solutions can provide the same solution to other governments(1), who is providing cyber security? (1), how long does it take to solve a problem? (1), CSP are reluctant to report incidents (breaches or outages) unless forced to do that (1), lack of transparency of CSP in reporting incidents (2) | 11.7 |
| **Ease of Use** | Ease of Use (4), easy to learn (4), ease to access (login), facilitate use of Government services for citizens(4), better user interfaces (2), improved usability(2) | 11 |
| **Social Norm** | large number of users sharing data on a common platform (1), use cases to understand main issues and how to solve them (1) | 1.4 |
| **PART 2 .** 1. "What are the most important perceived risks about using cloud computing services in your organization?" / 2. "Can you describe these risks and explain why they are so important?" | | |
| **CONCEPTS** | **PROPERTIES** (from Open coding) / **NUMBER OF TIMES** the property was mentioned | **%** |
| **Security Risk** | Loss of control (5),  privacy (3),  data security/protection (3), confidentiality (3), loss of availability(2),  loss of integrity(2), personnel security (3),  physical security(3), cyber security(1), cybercrime (1) | 37 |
| **Financial Risk** | Need to maintain the legacy systems(4), Locking into existing contracts, financial penalties from with withdrawing(4), new and complex procurement procedures (2), need to prove efficiency and effectiveness before spending money on new IT systems(2), Financial risk(2), High legal fees to personalise terms and conditions of use(1), Costs higher than expected(1) | 24.6 |
| **Social Risk** | Reputation could be damaged in case of security incidents…especially in case of leakage of personal data, services provided are not available/efficient, cost savings and IT efficiency is not confirmed (7), social risk (3), must be supported by Government to lower risk of reputation image (1), media can negatively influence public opinion after cyber incident (1) | 18.5 |
| **Performance Risk** | Performance risk (2),  cloud services vs legacy systems (more or less risk) (2), reliability (1),  need for a pilot test (1), system does not perform as advertised (1), Cloud services could not achieve expectations (1) | 12.3 |
| **Time risk** | Time to understand terms and conditions of use (1), compliance with storing personal data (1), time to research and implement new solutions(1), time to define and agree SLAs (1), time to learn (1) | 7.7 |
| **PART 3 .** 1. "How important is security risk about using cloud services within your organization and why?" / 2. "What kind of security risk are important to define the security risk of cloud-services?". | | |
| **CONCEPTS** | **PROPERTIES** (from Open coding) / **NUMBER OF TIMES** the property was mentioned | **%** |
| **Loss of service control** | if something happen we need quick answers (7) CSP can change rules without our consent (6) not marked information sharing risk (2) how to share sensitive  info (4) | 22.9 |
| **Data security/ protection** | where is located the CSP?(3)  Where are stored the data? (2)  who can access? (2) Are accesses  auditable?(1) are data protected from the data centre to the user? (1) UK law is very cautious about personal data and there are legal issues if data is not stored in the UK (3) Are logs accessible? (2) Are data securely stored  with adequate level of protection? (2) | 19.3 |
| **Cyber security** | cybercrime (1), cyber-attack (4), data leakage (1), leakage of sensitive information (1), concentration of data means lots of interests and threats (8) | 18.1 |
| **Personnel security** | who is the administrator of the system? (2) Who is providing technical support? (2) Authorizations policies and procedures (2)   control to access cloud data and infrastructure (malicious insiders) (3) how to ensure the need to know principle? (4) Only authorised people can access some data (1) | 16.9 |
| **Loss of privacy** | stolen pc with access to cloud services (13) | 15.7 |
| **Loss of confidentiality** | loss of data (4), stolen pc (2), human error (5) | 13.2 |
| **Loss of availability** | can not access your data if CSP is not working or reachable (5) outages (2), info availability to the right person (1) IT infrastructure is a strategic asset and if cloud is not available means to expand the risk (1) | 10.8 |
| **Loss of integrity** | loss of integrity (8) | 9.6 |
| **Physical security** | it is possible to access from difference location and devices (4) compliance to physical security standard (2) | 7.2 |
| **Lack of liability** | who is liable for poor quality of services provided ? (2) who is responsible for security incidents? (2) | 4.8 |
| **Identity management** | many user across different public sector organizations (3) | 3.6 |

# REFERENCES

[1] European Commission, "European Cloud Computing Strategy | Digital Agenda for Europe," 2012. [Online]. Available: http://ec.europa.eu/digital-agenda/en/european-cloud-computing-strategy. [Accessed: 23-Apr-2015].

[2] R. Kasper, "Perceptions of risk and their effects on decision making," Soc. Risk Assess., 1980.

[3] G. T. Gardner and L. C. Gould, "Public Perceptions of the Risks and Benefits of Technology '," vol. 9, no. 2, 1989.

[4] H. Otway and K. Thomas, "Reflections on Risk Perception and Policy1,2," Risk Anal., vol. 2, no. 2, pp. 69–82, Jun. 1982.

[5] S. Bellman, G. L. Lohse, and E. J. Johnson, "Predictors of online buying behavior," Commun. ACM, vol. 42, no. 12, pp. 32–38, 1999.

[6] D. Walker and F. Myrick, "Grounded theory: an exploration of process and procedure.," Qual. Health Res., vol. 16, no. 4, pp. 547–59, Apr. 2006.

[7] B. Zwattendorfer and K. Stranacher, "Cloud Computing in E-Government across Europe," Technol. Innov. Democr. Gov. Governance. Springer Berlin Heidelberg, 181-195., 2013.

[8] T. Haeberlen, D. Liveri, and M. Lakka, "Good Practice Guide for Securely Deploying Governmental Clouds," pp. 1–46, 2013.

[9] M. Miller, Cloud Computing: Web-Based Applications That Change the Way You Work and Collaborate Online. Que Publishing, 2008.

[10] R. L. Grossman, "The case for cloud computing," IT Prof., vol. 11, no. 2, pp. 23–27, 2009.

[11] M. Greer, "Software as a service inflection point: Using cloud computing to achieve business agility, iUniverse," Bloom., 2009.

[12] A. Vile and J. Liddle, TheSavvyGuideTo HPC, Grid, Data Grid, Virtualisation and Cloud Computing. 2008.

[13] J. W. S. Ali Khajeh-Hosseini, David Greenwood and I. Sommerville, "The Cloud Adoption Toolkit: supporting cloud adoption decisions in the enterprise," Softw. - Pract. Exp., vol. 42, no. 7, pp. 447–465, 2012.

[14] M. Armbrust, I. Stoica, M. Zaharia, A. Fox, R. Griffith, A. D. Joseph, R. Katz, A. Konwinski, G. Lee, D. Patterson, and A. Rabkin, "A view of cloud computing," Commun. ACM, vol. 53, no. 4, p. 50, Apr. 2010.

[15] K. K. Smitha, K. Chitharanjan, and T. Thomas, "Cloud based e-governance system : A survey," Procedia Eng., vol. 38, pp. 3816–3823, 2012.

[16] S. Paquette, P. T. Jaeger, and S. C. Wilson, "Identifying the security risks associated with governmental use of cloud computing," Gov. Inf. Q., vol. 27, no. 3, pp. 245–253, 2010.

[17] Q. Zhang, L. Cheng, and R. Boutaba, "Cloud computing: state-of-the-art and research challenges," J. Internet Serv. Appl., vol. 1, no. 1, pp. 7–18, Apr. 2010.

[18] N. Leavitt, "Is cloud computing really ready for prime time?," Computer (Long. Beach. Calif)., 2009.

[19] A. Tripathi and B. Parihar, "E-Governance challenges and cloud benefits," Proc. - 2011 IEEE Int. Conf. Comput. Sci. Autom. Eng. CSAE 2011, vol. 1, pp. 351–354, 2011.

[20] A. Lin and N.-C. Chen, "Cloud computing as an innovation: Percepetion, attitude, and adoption," Int. J. Inf. Manage., pp. 1–8, Apr. 2012.

[21] Barry Bozeman, "Culture in Public and Private Organizafions," vol. 58, no. 2, pp. 109–118, 1998.

[22] S. B. SITKIN and L. R. WEINGART, "DETERMINANTS OF RISKY DECISION-MAKING BEHAVIOR: A TEST OF THE MEDIATING ROLE OF RISK PERCEPTIONS AND PROPENSITY.," Acad. Manag. J., vol. 38, no. 6, pp. 1573–1592, Dec. 1995.

[23] A. Fiegenbaum and H. Thomas, "ATTITUDES TOWARD RISK AND THE RISK-RETURN PARADOX: PROSPECT THEORY EXPLANATIONS.," Acad. Manag. J., vol. 31, no. 1, pp. 85–106, Mar. 1988.

[24] S. Jackson and J. Dutton, "Discerning threats and opportunities," Adm. Sci. Q., 1988.

[25] K. R. MacCrimmon and D. A. Wehrung, "Characteristics of Risk Taking Executives," Manage. Sci., vol. 36, no. 4, pp. 422–435, Apr. 1990.

[26] A. Gore, From red tape to results: Creating a government that works better & costs less: Report of the National Performance Review. 1993.

[27] R. Osborn and D. Jackson, "Leaders, riverboat gamblers, or purposeful unintended consequences in the management of complex, dangerous technologies," Acad. Manag. J., 1988.

[28] H. Rainey, S. Pandey, and B. Bozeman, "Research note: Public and private managers' perceptions of red tape," Public Adm. Rev., 1995.

[29] H. Rainey and B. Bozeman, "Comparing public and private organizations: Empirical research and the power of the a priori," J. public Adm. Res. …, 2000.

[30] S. Pandey and P. Scott, "Red tape: A review and assessment of concepts and measures," J. Public Adm. Res. Theory, 2002.

[31] M. G. Morgan, M. Henrion, and M. Small, Uncertainty: A Guide to Dealing with Uncertainty in Quantitative Risk and Policy Analysis. Cambridge University Press, 1992.

[32] D. Grewal, J. Gotlieb, and H. Marmorstein, "The moderating effects of message framing and source credibility on the price-perceived risk relationship," J. Consum. Res., 1994.

[33] S. Cunningham, "The major dimensions of perceived risk," Risk Tak. Inf. Handl. Consum. Behav. 82-108., 1967.

[34] L. Kaplan, G. Szybillo, and J. Jacoby, "Components of perceived risk in product purchase: A cross-validation.," J. Appl. Psychol., 1974.

[35] A. Benlian and T. Hess, "Opportunities and risks of software-as-a-service: Findings from a survey of IT executives," Decis. Support Syst., vol. 52, no. 1, pp. 232–246, Dec. 2011.

[36] W. Wu, L. Lan, and Y. Lee, "Exploring decisive factors affecting an organization's SaaS adoption: A case study," Int. J. Inf. Manage., 2011.

[37] M. Janssen and A. Joha, "Challenges for Adopting Cloud-Based Software as a Service (SaaS) in the Public Sector," Proc. Eur. Conf. Informait. Syst. (ECIS 2011), 2011.

[38] A. Strauss and J. Corbin, "Basics of qualitative research: Procedures and techniques for developing grounded theory," ed Thousand Oaks, CA Sage, 1998.

[39] A. Lee and R. Baskerville, "Generalizing generalizability in information systems research," Inf. Syst. Res., 2003.

[40] B. Glaser and A. Strauss, "The discovery grounded theory: strategies for qualitative inquiry," London, Engl. Wiedenfeld Nicholson, 1967.

[41] M. Larkin, S. Watts, and E. Clifton, "Giving voice and making sense in interpretative phenomenological analysis," Qual. Res. Psychol., 2006.

[42] S. Bird, D. Day, and J. Garofolo, "ATLAS: A flexible and extensible architecture for linguistic annotation," arXiv Prepr. cs/ …, 2000.

[43] H. Heath and S. Cowley, "Developing a grounded theory approach: a comparison of Glaser and Strauss," Int. J. Nurs. Stud., vol. 41, no. 2, pp. 141–150, Feb. 2004.

[44] A. Benlian and T. Hess, "Opportunities and risks of software-as-a-service: Findings from a survey of IT executives," Decis. Support Syst., vol. 52, no. 1, pp. 232–246, 2011.

[45] W. W. Wu, "Developing an explorative model for SaaS adoption," Expert Syst. Appl., vol. 38, no. 12, pp. 15057–15064, 2011.

[46] C. Czosseck, R. Ottis, and A. Taliharm, "Estonia after the 2007 cyber attacks: Legal, strategic and organisational changes in cyber security," Case Stud. Inf. …, 2013.

[47] Marnix Dekker, "Security Framework for Governmental Clouds — ENISA." 2015.

[48] M. Siegrist, C. Keller, H. Kastenholz, S. Frey, and A. Wiek, "Laypeople's and experts' perception of nanotechnology hazards.," Risk Anal., vol. 27, no. 1, pp. 59–69, Feb. 2007.

[49] L. Rittenberg and F. Martens, "COSO: Understanding and Communicating Risk Appetite," Comm. Spons. Organ. Treadw. Comm., pp. 1–32, 2012.

## AUTHORS

**Gianfranco Elena** is a doctoral researcher at Glasgow University. He has been working as CTO and CIO in the Italian Army and NATO organizations for more than 20 years. His expertise is in cyber security and risk assessment of cloud computing services. His research contributes to improve the decision making process about the adoption of Cloud computing in Government organizations.

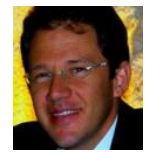

**Professor Chris Johnson** is Head of Computing Science at Glasgow University. His research increases the resilience of critical infrastructures. He is a software specialist on the SESAR scientific board advising the European Commission on the future of Air Traffic Management. He also focuses on the interactions between safety and security - for example, developing techniques so that we can safely close down a civil nuclear reactor even after malware has been detected.

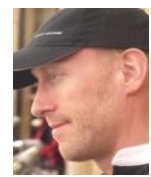